\documentclass[12pt,draftclsnofoot,onecolumn]{IEEEtran}
\usepackage[noadjust]{cite}
\usepackage[subsection]{algorithm}
\usepackage{amsmath,amssymb,amsbsy,algorithmic,bm,url,color}
\ifCLASSINFOpdf
   \usepackage[pdftex]{graphicx}
   \DeclareGraphicsExtensions{.pdf,.jpeg,.png}
\else
   \usepackage[dvips]{graphicx}
   \DeclareGraphicsExtensions{.eps}
\fi

\def\Ibf{{\mathbf I}}

\newcommand{\Tr}{\mathrm{Tr}}
\newcommand{\rank}{\mathrm{rank}}

\newcommand{\W}{\mathbf{W}}

\newcommand{\Pwrcon}{\Omega_{+}(P,\xi_p)}

\begin{document}
\title{Prescient Precoding in Heterogeneous DSA Networks with Both Underlay and Interweave MIMO Cognitive Radios}
\author{Amitav~Mukherjee and A.~Lee~Swindlehurst
\thanks{The authors are with the Dept.~of Electrical Engineering and Computer Science,
University of California at Irvine, CA 92697-2625, USA. {\tt (e-mail: \{amukherj; swindle\}@uci.edu)}}
\thanks{This work was supported by the National Science Foundation under grant CCF-0916073.}
}
\maketitle

\begin{abstract}
This work examines a novel heterogeneous dynamic spectrum access network where the primary users (PUs) coexist with both underlay and interweave cognitive radios (ICRs); all terminals being potentially equipped with multiple antennas. Underlay cognitive transmitters (UCTs) are allowed to transmit concurrently with PUs subject to interference constraints, while the ICRs employ spectrum sensing and are permitted to access the shared spectrum only when both PUs and UCTs are absent. We investigate the design of MIMO precoding algorithms for the UCT that increase the detection probability at the ICRs, while simultaneously meeting a desired Quality-of-Service target to the underlay cognitive receivers (UCRs) and constraining interference leaked to PUs. The objective of such a proactive approach, referred to as \emph{prescient} precoding, is to minimize the probability of interference from ICRs to the UCRs and primary receivers due to imperfect spectrum sensing. We begin with downlink prescient precoding algorithms for multiple single-antenna UCRs and multi-antenna PUs/ICRs. 
We then present prescient block-diagonalization algorithms for the MIMO underlay downlink where spatial multiplexing is performed for a plurality of multi-antenna UCRs.
Numerical experiments demonstrate that prescient precoding by UCTs provides a pronounced performance gain compared to conventional underlay precoding strategies.
\end{abstract}


\section{INTRODUCTION}\label{sec:intro}
Dynamic spectrum access (DSA) is emerging as a promising solution to enable better
utilization of the radio spectrum, especially in bands that are currently
under-utilized \cite{Niyato_Han}. DSA partitions wireless terminals into categories of primary (licensed) and secondary (cognitive radio) users, where the primary users have priority in accessing the shared spectrum.  Furthermore, the two most prevalent classifications of secondary users are underlay cognitive radios and interweave cognitive radios (ICRs), following the terminology of \cite{Jafar09}. The underlay paradigm mandates that concurrent secondary
and primary transmissions may occur only if the interference generated by the underlay cognitive transmitters (UCTs) at the primary receivers (PRs) is below some acceptable threshold. In contrast, ICRs are allowed to opportunistically use the spectrum only when it is not occupied by a primary transmitter (PT) with priority. In the absence of standard control channels or coordinated medium access between the primary and secondary users, the ICRs must periodically sense the spectrum for the presence of PTs \cite{Niyato_Han,Jafar09,SeguraWang10} and cease transmission upon detection. Inevitably, imperfect ICR spectrum sensing due to channel fading and other impairments will lead to unintentional interference at the underlay cognitive receivers (UCRs) and PRs.

Underlay and ICR networks have been studied separately in extensive detail for both single-antenna and multiple-input-multiple-output (MIMO) terminals \cite{Niyato_Han,Jafar09}.
The use of multiple antennas in ICRs has been suggested for improved spectrum sensing capabilities by means of receive diversity \cite{Scharf06}-\cite{Vilar11}. MIMO systems have also been investigated in the context of underlay DSA networks, where multiple transmit antennas are used by UCTs for beamforming and to control the interference to the PRs, assuming either complete or partial channel state information (CSI) at the SU transmitter \cite{Zhang08}-\cite{Phan09}. However, there is little if any prior work that examines heterogeneous DSA networks with \emph{both} UCTs and ICRs attempting to coexist simultaneously with primary users. Note that such a scenario is significantly different from hybrid secondary users that are capable of both underlay and interweave cognition \cite{Yanikomeroglu10,Chakravarthy09}.

Therefore, this work examines a fundamentally novel heterogeneous DSA network where the primary users share their spectrum with both UCRs and ICRs; all terminals being potentially equipped with multiple antennas. Specifically, we investigate the design of MIMO precoding algorithms for a underlay downlink network with multiple UCRs and interweave radios. The heterogeneous DSA network presents a myriad of conflicting objectives for the underlay transmitter, since it must mitigate the multi-user interference among its own UCRs, constrain the interference leaked to PRs, and ensure that the detection probability of the ICRs is high so as to preemptively avoid interference from them. Consequently, this paper is devoted to the design of novel precoding algorithms, collectively referred to as \emph{prescient} precoding, that balance these competing objectives. The aim of prescient precoding is to reduce the probability of interference due to imperfect spectrum sensing from ICRs to the underlay and primary receivers, while simultaneously meeting their QoS/interference temperature requirements. 
Therefore, our contributions include the following:
\begin{itemize}
\item We introduce a novel heterogeneous DSA network with both underlay and interweave radios, and demonstrate that existing underlay precoding schemes are highly suboptimal for such networks.
\item We design new prescient downlink precoding schemes of varying complexity for multiple single-antenna UCRs and MIMO PRs/ICRs.
\item We present prescient block-diagonalization algorithms for the MIMO UCR downlink channel when all UCRs are equipped with multiple antennas in addition to the PRs and ICRs.
\item We demonstrate via numerical simulations that diverting resources from the UCRs to increase the detection probability at the ICRs can significantly suppress unintentional ICR interference.
\end{itemize}

The paper is organized as follows. Section~\ref{sec:model} introduces the mathematical model of the DSA network and the spectrum sensing performance of the ICRs. Prescient downlink precoding algorithms for the case of single-antenna underlay receivers are proposed in Section~\ref{sec:PDP}. 
 Section~\ref{sec:MIMOUnderlayRxs} outlines a prescient block-diagonalization algorithm for a MIMO downlink channel with multi-antenna underlay receivers. Selected numerical examples are shown in Section~\ref{sec:Sim}, and we conclude in Section~\ref{sec:Conclusion}.

\emph{Notation}:
We will use $\mathcal{CN}(\mathbf{0},\mathbf{Z})$ to denote a circularly symmetric complex
Gaussian distribution with zero mean and covariance matrix $\mathbf{Z}$,
$\mathcal{E}\{\cdot\}$ to denote expectation, $\text{vec}(\cdot)$ the matrix column stacking operator, $(\cdot)^T$ the transpose, $(\cdot)^H$ the Hermitian
transpose, $(\cdot)^{-1}$ the matrix inverse, $\Tr(\cdot)$ the
trace operator, $\left| \cdot \right|$ the matrix determinant, and
$\mathbf{I}$ an identity matrix of appropriate dimension.

\section{Mathematical Model}\label{sec:model}

\subsection{Signal and Network Model}
\begin{figure}[htbp]
\centering
\includegraphics[width=\linewidth]{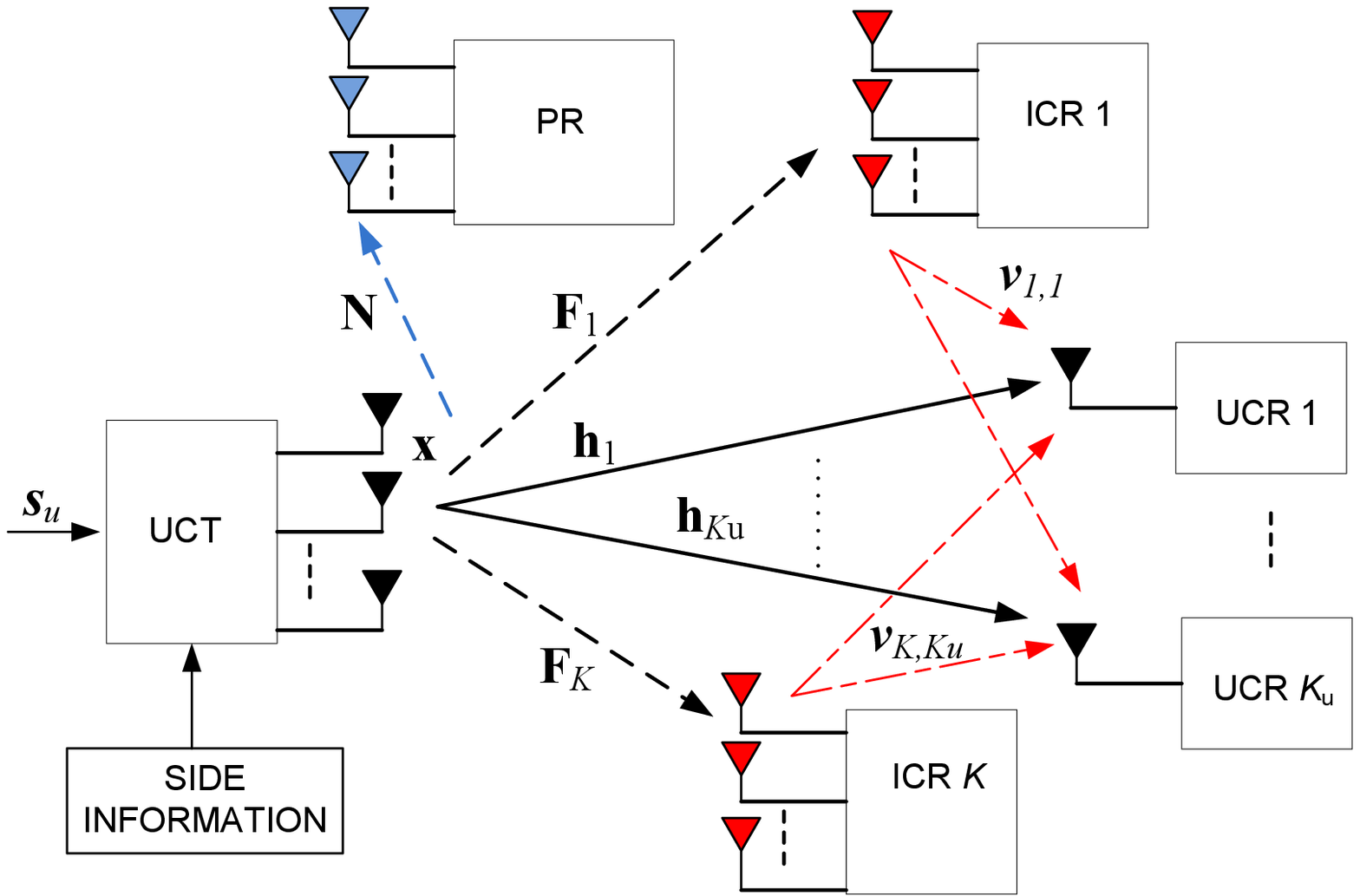}
\caption{Cognitive radio network with a multi-antenna underlay transmitter, $K_u$ underlay receivers, a single MIMO primary receiver, and $K$ spectrum-sensing multi-antenna interweave cognitive radios. The primary transmitter and ICR-to-PR interfering links are not shown for clarity.}
\label{fig_network}
\end{figure}

Consider a downlink network with a $t_u$-antenna UCT, $K_u$ single-antenna UCRs as its intended destinations where $K_u \leq t_u$, $K$ multi-antenna ICRs with $r_I$ antennas each, and a single PT-PR pair with $t_p$ and $r_p$ antennas respectively, as depicted in Fig.~\ref{fig_network}. 
The scenario with multi-antenna UCRs is presented in Sec.~\ref{sec:MIMOUnderlayRxs}. Multiple PRs can be accommodated in the current model by aggregating them into a single virtual PR.

Assuming linear precoding, the UCT downlink transmit signal at time index $t$ is written as
\begin{equation}\label{eq:MultiPU_x}
{\mathbf{x}}\left(t\right) = \sum\limits_{k = 1}^{K_u } {{\mathbf{w}}_i s_{u,i}\left(t\right)  =  {\mathbf{W}}}{\mathbf{s}}_u\left(t\right),
\end{equation}
where $\mathbf{W} \in \mathbb{C}^{t_u \times K_u }=\left[ {\begin{array}{*{20}{c}}
  {{{\mathbf{w}}_1}}& \ldots &{{{\mathbf{w}}_{{K_u}}}}
\end{array}} \right]$ is the precoding matrix whose columns represent individual beamforming vectors, and ${\mathbf{s}}_u\left(t\right)  \in \mathbb{C}^{K_u  \times 1}$ is the collection of i.i.d. underlay information symbols
 drawn from an $M$-ary phase-shift keying constellation
with second-order statistics $E\left\{ {{{\mathbf{s}}_u}{\mathbf{s}}_u^H} \right\} =~{\mathbf{I}}$. A power constraint $\Tr\left({\W\W^H}\right)\le P$ is imposed on the UCT signal. Furthermore, the UCT designs its transmit signal so as to ensure that the detection probability at the ICRs is satisfactory and the interference temperature at the PR remains below a pre-specified threshold ${\xi _p}$, as explained in Sec.~\ref{sec:PDP}. 

Suppressing the time index, the received signal at UCR $k$ in the \emph{absence} of ICR interference (i.e., with perfect spectrum sensing) is
\begin{equation}
y_k={\mathbf{h}}_k\W \mathbf{s}_u + n_k
\end{equation}
where 
${\mathbf{h}}_k\in \mathbb{C}^{1 \times t_u}$ is the corresponding complex channel vector from the UCT, and $n_k$ is a circularly symmetric zero-mean complex Gaussian noise sample with variance $\sigma^2_k$ which includes interference from the PT.

We list below the major assumptions regarding the heterogeneous DSA network.
\begin{itemize}
\item We assume a \emph{partial CSI} model at the UCT,
which is defined to mean that the UCT always has knowledge of the
instantaneous realizations of all the downlink channels ($\{\mathbf{h}_k\}^{K_u}_{k=1}$) and UCT-PR
($\left\{\mathbf{h}_k\right\}_{k=1}^K$) channels, but may know only the distribution of its channels to the ICRs and
the ICR-to-UCR channels. 
\item The UCT has knowledge of the ICR transmit powers and the parameters of the spectrum sensing scheme deployed at the ICRs, which in practice are likely to be pre-defined by spectrum regulatory agencies.
\item There is no coordination between the UCT and ICRs. The UCT and PRs have a limited coordination in order to exchange CSI and tolerable interference limits.
\item All ICRs are half-duplex, which precludes for example simultaneous data transmission and spectrum sensing.
 We only consider in-band spectrum sensing; i.e., sensing is conducted on the same band that is used for data transmission. 
\item The UCRs employ single-user decoding and treat all ICR/PT interference as noise. The interference from the ICRs is assumed to be instantaneous, i.e., the processing delay due to spectrum sensing is neglected.
\end{itemize}

\subsection{ICR Spectrum Sensing}\label{sec:NoncoopSensing}
We assume that prior to transmitting, each ICR uses a binary hypothesis test based on $\tilde{M} r_I$ spatio-temporal samples to determine whether or not the band is occupied.  Under the two hypotheses, the signal received by the $i^{th}$ ICR at time $n$ is: 
\begin{subequations}
\begin{align}
   {\mathcal{H}_0:} & \:{{{\mathbf{z}}_i}\left[ n \right] = {{\mathbf{m}}_i}\left[ n \right],} & {n = 0,\ldots ,\tilde
   M-1}\label{eq:HypoTestH0}\\
{\mathcal{H}_1:} & \:{{{\mathbf{z}}_i}\left[ n \right] = {{\mathbf{q}}_i}\left[ n \right] + {{\mathbf{m}}_i}\left[ n \right],} & {n = 0,\ldots ,\tilde M-1}
   \label{eq:HypoTestH1}
 \end{align}
\end{subequations}
where $\mathbf{m}_{i}[n]\sim\mathcal{CN}(\mathbf{0},\epsilon_i^2\Ibf)$ is temporally
uncorrelated background Gaussian noise of known variance and
$\mathbf{q}_{i}[n]$ represents the presence of a signal in the band.  The
$\tilde M r_I$ complex samples are composed of $2\tilde M r_I$ independent
real and imaginary components \cite{Li08}.  We are only interested in
the case where $\mathbf{q}_{i}[n]=\{q_{i,1}[n],\ldots,q_{i,r_I}[n]\}$ is due to the UCT and also possibly the PT,
in which case
\begin{equation}
q_{i,j}[n] = {\mathbf{f}}_{i,j} {\mathbf{Ws}_u\left[ n \right]}
   +{\mathbf{d}}_{i,j} {\mathbf{s}_p\left[ n \right]} \; , j=1,\ldots,r_I,
\end{equation}
where channels ${\mathbf{f}}_{i,j}\in \mathbb{C}^{1 \times t_u}$ from
the UCT and ${\mathbf{d}}_{i,j}\in \mathbb{C}^{1 \times t_p}$ from the
PT are assumed to be invariant over the $\tilde M$ samples, and
$\mathbf{s}_p\in \mathbb{C}^{t_p \times 1}$ is the PT transmit signal
with total power $P_t$.


A broad range of spectrum sensing algorithms with varying levels of complexity and requisite \emph{a priori} information have been proposed in the literature \cite{Scharf06}-\cite{Vilar11}. On one hand, the optimal matched-filter detector has the most prohibitive requirements for CSI and PT signal information, while on the other hand non-coherent energy detection is the simplest possible detector since it only requires an accurate estimate of the noise variance $\epsilon_i^2$.  A range of composite generalized likelihood ratio tests (GLRT) and feature detectors lie in between these extremes. For this work, we assume the ICRs employ non-coherent energy detection due to its simplicity and the fact that it is unnecessary to distinguish between the UCT and PT signals.

The test statistic and threshold test for the energy detector is given by \cite{Hong09}
\begin{equation}\label{eq:test_statistic}
T_i  = \sum\limits_{n = 0}^{\tilde M - 1} {\sum\limits_{j = 1}^{{r_I}} {{{\left| {{z_{i,j}}\left[ n \right]} \right|}^2}} }; \quad T_i \mathop  \gtrless \limits_{{\mathcal{H}_0}}^{{\mathcal{H}_1}} \lambda_i
\end{equation}
where $\lambda_i$ is the detection threshold. We begin our development by analyzing the detection probability $P_{D,i}$ at ICR $i$ assuming deterministic channels and signals
from the UCT and PT. 
Under the null hypothesis $\mathcal{H}_0$, we see from (\ref{eq:HypoTestH0}) that ${z}_{i,j} \left[ n \right]  \sim \mathcal{CN}\left( {0,\epsilon_i^2 } \right)$, whereas under the alternative hypothesis $\mathcal{H}_1$ we have ${z}_{i,j} \left[ n \right]  \sim \mathcal{CN}\left( \mu_{i,j}\left[ n \right]={\mathbf{f}}_{i,j} {\mathbf{Ws}_u\left[ n \right]} +{\mathbf{d}}_{i,j} {\mathbf{s}_p\left[ n \right]},\epsilon_i^2 \right)$.
Therefore, the test statistic $T_i$ is the sum of the squares of $2\tilde M r_I$ independent real 
Gaussian random variables and thus follows a $\chi$-squared distribution
under both hypotheses:
\begin{equation}
\begin{array}{*{20}c}
   T_i & { \sim {\frac{{\epsilon_i^2 }}{2}} \chi _{2\tilde M r_I}^2 } & {{\text{under }}\mathcal{H}_0 }  \\
T_i & { \sim {\frac{{\epsilon_i^2 }}{2}} \chi_{2\tilde M r_I}^{'2}\left( \rho  \right)} & {{\text{under }}\mathcal{H}_1 }  \\
 \end{array}
\end{equation}
where the noncentrality parameter $\rho  = \epsilon _i^{ - 2}{\sum\nolimits_{n = 0}^{\tilde M - 1} {\sum\nolimits_{j = 1}^{{r_I}} {\left| {{\mu _{i,j}}\left[ n \right]} \right|} } ^2}$ is a function of $\mathbf{W}$.

Since we have an even number of samples $2\tilde M r_I$ (real and imaginary components of each sample), the false alarm probability follows immediately from the central chi-square cdf as \cite{Kay}
\begin{equation}\label{eq:PFAk}
P_{FA,i}  = e^{ - \frac{\lambda_i }
{{\epsilon_i^2 }}} \sum\limits_{r = 0}^{\tilde Mr_I-1 } {\frac{1}
{{r!}}\left( {\frac{\lambda_i }
{{\epsilon_i^2 }}} \right)} ^r \; ,
\end{equation}
where $\lambda_i  = \epsilon_i^2 Q_{\chi _{2\tilde M r_I}^2 }^{ - 1} \left( {P_{f} } \right)$ is chosen to satisfy a target false alarm rate $P_{f}$, and $Q_{\chi _{2\tilde M r_I}^2 } \left(  \cdot  \right)$ is the complementary cdf of the central chi-square distribution. The detection probability is given by
\begin{equation}\label{eq:P_Dk}
P_{D,i}={Q_{\tilde M{r_I}}}\left( {\sqrt {\rho ,} \sqrt {\frac{{2{\lambda _i}}}{{\epsilon _i^2}}} } \right),
\end{equation}
 where $Q_{\tilde M{r_I}} \left(  \cdot,\cdot  \right)$ is the generalized Marcum $Q$-function \cite{Kay07}.
 As the number of samples $\tilde M$ grows large, $T_i$ approaches a Gaussian random variable in distribution by the central limit theorem (CLT). Under hypothesis $\mathcal{H}_1$ the CLT yields $T_i\sim \mathcal{N}\left( {\epsilon_i^2\tilde M r_I+\rho, \epsilon_i^4\tilde M r_I+2\epsilon_i^2\rho} \right)$ and the corresponding detection probability
 \begin{equation}\label{eq:P_DkCLT}
P_{D,i} \simeq Q\left(  \frac{\lambda_i-\epsilon_i^2\tilde M r_I-\rho}{\epsilon_i \sqrt{\epsilon_i^2\tilde M r_I+2\rho}}  \right),
\end{equation}
where $Q\left(  \cdot  \right)$ is the Gaussian $Q$-function.

\subsection{ICR Performance Prediction at UCT}
The ability of the UCT to predict the spectrum-sensing performance of the ICRs is an important ingredient of the prescient precoding paradigm. Under the partial CSI assumption, it is highly unlikely that the UCT has knowledge of the PT-to-ICR channel realizations and signals required to compute (\ref{eq:P_Dk}) or (\ref{eq:P_DkCLT}). A more plausible scenario is that the UCT knows the realizations of its channels $\{{\mathbf{F}}_{i}\}$ to the ICRs, and assumes the PT-to-ICR channels undergo Rayleigh fading with distribution ${{\mathbf{d}}_{i,j}}\sim \mathcal{CN}\left( {{\mathbf{0}},\sigma _{d,i}^2{\mathbf{I}}} \right)$ $\forall i,j$.

Going one step further, the UCT may not have knowledge of the instantaneous realizations of its channels to the ICRs either. In order to gauge the energy detection performance of the ICRs, the UCT assumes a Rayleigh fading scenario such that ${{\mathbf{f}}_{i,j}}\sim \mathcal{CN}\left( {\mathbf{0},\sigma _{f,i}^2{\mathbf{I}}} \right)$ $\forall i,j$, and ${{\mathbf{d}}_{i,j}}\sim \mathcal{CN}\left( {{\mathbf{0}},\sigma _{d,i}^2{\mathbf{I}}} \right)$  as before. Furthermore, the UCT and PT signals are each assumed to be drawn with uniform probability from a complex $M$-ary PSK constellation, and all channels, signals, and AWGN samples are mutually independent. Given these assumptions, the ICR samples ${z}_{i,j} \left[ n \right]$ are distributed as independent Gaussian random variables \cite{Hong09} for both hypotheses. The false-alarm rate is clearly identical to that in (\ref{eq:PFAk}) since it is channel-independent. Under $\mathcal{H}_1$,  $E\left\{ {{z_{i,j}}\left[ n \right]} \right\} = 0$ and $  \sigma _{z,i}^2\triangleq \operatorname{var} \left\{ {{z_{i,j}}\left[ n \right]} \right\}= 2\sigma _{f,i}^2\Tr\left( {{\mathbf{W}}{{\mathbf{W}}^H}} \right) + 2{P_t}{t_p}\sigma _{d,i}^2 + \varepsilon _i^2$.
Thus, $T_i{ \sim {\frac{{\sigma _{z,i}^2 }}{2}} \chi _{2\tilde M r_I}^2 }$ and the corresponding average detection probability is
\begin{equation}\label{eq:barPDi}
\bar P_{D,i}  = e^{ - \frac{\lambda_i }
{{\sigma _{z,i}^2 }}} \sum\limits_{r = 0}^{\tilde Mr_I-1 } {\frac{1}
{{r!}}\left( {\frac{\lambda_i }
{{\sigma _{z,i}^2 }}} \right)} ^r.
\end{equation}

From the UCT's perspective, a missed detection (Type II error) at any
of the ICRs leads to interference at the underlay receivers, and this
phenomenon plays a pivotal role in the prescient precoding principle.
It will be useful to define the Bernoulli-distributed indicator
function $F_i$ as
\begin{equation}\label{eq:F_k}
F_i  = \left\{ {\begin{array}{*{20}c}
   1 & {{\text{with probability }}\left( {1 - P_{D,i}  } \right)}  \\
   0 & {{\text{with probability }}\left( {P_{D,i} } \right)}.  \\
 \end{array} } \right.
\end{equation}
$F_i$ therefore models the likelihood that ICR $i$ unintentionally causes interference to the underlay and primary receivers, and is a function of $\mathbf{W}$ via $P_{D,i}$.

 Clearly, it is in the UCT's best interest to ensure that the probability of missed detection at the ICRs is made as small as possible, or equivalently, that the probability of detection is made as large as possible. To this end, we introduce the paradigm of prescient precoding in the next section in order to improve the reliability of the underlay downlink.

\section{Prescient Downlink Precoding}\label{sec:PDP}
It has been elegantly established that the capacity region of a conventional non-cognitive multi-antenna downlink channel without structured interference is achieved through non-linear dirty-paper coding, since all transmitted signals are known non-causally to the transmitter \cite{Weingarten06}. However, linear precoding schemes for the multiuser downlink have been extensively studied due to their significantly lower complexity and near-capacity performance in certain regimes, and thus we focus on linear transmit preprocessing at the UCT. By definition, the UCT must limit the (instantaneous or average) interference it causes to the PR to a predefined threshold $\xi_p$:
\begin{equation}
\Tr\left( {{\mathbf{NW}}{{\mathbf{W}}^H}{{\mathbf{N}}^H}} \right) \leq {\xi _p}
\end{equation}
if the instantaneous channel $\mathbf{N}\in \mathbb{C}^{r_p \times t_u}$ to the PR is known.

The signal at an arbitrary UCR inclusive of ICR interference due to missed detections can be written as
\begin{equation}
y_k  = {\mathbf{h}}_k {\mathbf{w}}_k s_{u,k}  + \underbrace{\sum\limits_{j \ne k}^{K_u } {{\mathbf{h}}_k {\mathbf{w}}_j s_{u,j} }}_{{\text{intra - UCR interference}}}  + \underbrace{\sum\limits_{i = 1}^K {F_i \mathbf{v}_{k,i} \mathbf{s}_{I,i} }}_{\text{ICR interference}} {+}\: n_k, \quad k=1,\ldots,K_u,
\end{equation}
where $\mathbf{v}_{k,i}\sim \mathcal{CN}\left(\mathbf{0},\sigma_{v,i}^2\mathbf{I}\right)$ and $\mathbf{s}_{I,i}\in \mathbb{C}^{r_I \times 1}$ represent the $({1 \times r_I})$ interfering channel and signal vector of power $P_i$ from ICR $i$.
We are interested in the characteristics of the aggregate ICR interference power at the $k^{th}$ UCR, defined as
\begin{equation}\label{eq:I0}
I_{k}\left(\mathbf{W}\right)  = \sum\limits_{i = 1}^K {F_i \left\| {\mathbf{v}_{k,i} } \right\|^2 P_i }.
\end{equation}
Taking the expectation of the ICR interference power
in~(\ref{eq:I0}) with respect to indicator functions $\left\{ {F_i } \right\}_{i = 1}^K$ and the ICR-UCR channels
$\left\{\mathbf{v}_{k,i}\right\}_{i=1}^K$ yields
\begin{equation}
{\bar I}_{k} \left( {\mathbf{W}} \right) = \sum\limits_{i = 1}^K {\left( {1 - P_{D,i}} \right)P_i r_I {\sigma^2_{v,i} }   }.
\end{equation}
The UCR SINR that can be computed at the UCT is then approximated as
\begin{equation}\label{eq:UCR_SINR}
\gamma_{k}  = \frac{{\left| {{\mathbf{h}}_k {\mathbf{w}}_k } \right|^2}}
{{\sum\nolimits_{j \ne k}^{K_u} {\left| {{\mathbf{h}}_k {\mathbf{w}}_j } \right|^2 }  + {\bar I}_{k}\left(\mathbf{W}\right)  + \sigma _k^2 }}, \quad k=1,\ldots,K_u,
\end{equation}
where the aggregate ICR interference ${\bar I}_{k}\left(\mathbf{W}\right)$ is a function of $\mathbf{W}$ via the spectrum-sensing detection probabilities.

In the remainder of this section, we present several prescient
design solutions for $\mathbf{W}$ that provide a tradeoff between
complexity and underlay downlink performance. The attribute of ``prescience'' derives from the fact that the UCT anticipates interference at the PR from SUs due to imperfect spectrum sensing and takes preemptive measures to avoid the same.

\subsection{Direct UCR Sum Rate Maximization}
 A wide variety of choices for $\mathbf{W}$ for conventional non-cognitive and underlay-only downlink channels have been explored in the literature. For example, a na\"{\i}ve transmission scheme that disregards ICR CSI and PR interference would be to apply a modified regularized channel inversion (RCI) precoder \cite{Swindlehurst05}, with
\begin{equation}
\label{eq:RCI}
{\mathbf{W}}_{CI} = \frac{1}
{{\sqrt \zeta  }}{\mathbf{H}}_{u}^H \left( {{\mathbf{H}}_{u} {\mathbf{H}}_{u}^H  +
\frac{K_u}{P} {\mathbf{I}}} \right)^{ - 1}
\end{equation}
where ${\mathbf{H}}_u  \triangleq \left[ {\begin{array}{*{20}c}
   {{\mathbf{h}}_1^T } &  \ldots  & {{\mathbf{h}}_{K_u}^T }  \\
 \end{array} } \right]^T$, given the scale factor $\zeta$ which is chosen as the smaller of the two scaling factors required to preserve the UCT transmit power and PR interference temperature constraints. However, the na\"{\i}ve RCI precoder does not account for the potential ICR interference $\bar{I}_k$, which can severely degrade the underlay sum-rate performance when $\bar{I}_k$ is the dominant term of the denominator in (\ref{eq:UCR_SINR}).
A more efficient usage of the side information available to the UCT is a direct sum-rate maximization approach that exploits knowledge of the ICR channels:
\begin{subequations}\label{eq:DirectRsMax}
\begin{align}
\mathop {\max }\limits_{\mathbf{W}} & \sum\limits_{k = 1}^{K_u} {\log _2 \left( {1 + \gamma_{k} } \right)}\\
\mathrm{s.t.} & \quad \Tr\left( {{\mathbf{NW}}{{\mathbf{W}}^H}{{\mathbf{N}}^H}} \right) \leq {\xi _p}\\
{}& \quad \Tr\left({{\mathbf{W}}{{\mathbf{W}}^H}} \right) \leq P.
\end{align}
\end{subequations}

The above problem is novel since the co-channel ICR interference term in the SINR is a function of the transmit signal itself. This is in sharp contrast with conventional single-cell \cite{Utschick10}, multi-cell \cite{Venturino10}, and underlay-only \cite{Zhang08}-\cite{Phan09} downlink beamforming problems where the co-channel interference is inevitably modeled as independent noise. While
signal-dependent interference is a well-studied problem in radar
signal processing \cite{Pillai00}, in our case this
dependence manifests itself in a much more complicated and non-linear
fashion involving exponential terms. We are faced with a non-convex objective function with multiple non-linear constraints, and at this point an analytical solution for $\mathbf{W}$ therefore appears to be intractable.

To solve the sum-rate maximization problem numerically, a gradient projection (GP) algorithm can be used, which will converge to at least a locally-optimal stationary point.
To compute the gradient of the UCR sum rate, we define the leakage term
\begin{equation}
{L_{k,j}} = \sum\limits_{j \ne k} {{{\left| {{{\mathbf{h}}_k}{{\mathbf{w}}_j}} \right|}^2}}  + {{\bar I}_k}\left( {\mathbf{W}} \right) + \sigma _k^2,
\end{equation}
and compute $\nabla _{\mathbf{W}} \left( R_s \right) = \left[ {\begin{array}{*{20}c}
   {\nabla^T_{{\mathbf{w}}_1 } \left( R_s \right)} &  \ldots  & {\nabla^T_{{\mathbf{w}}_{K_u } } \left( R_s \right)}  \\
 \end{array} } \right]^T$
 where
 \begin{align}
 \begin{split}
  {\nabla _{{{\mathbf{w}}_k}}}\left( {{R_s}} \right) &= \frac{1}{{\ln 2}}{\left( {1 + \frac{{{{\left| {{{\mathbf{h}}_k}{{\mathbf{w}}_k}} \right|}^2}}}{{{L_{k,j}}}}} \right)^{ - 1}}\frac{{2{\mathbf{h}}_k^H{{\mathbf{h}}_k}{{\mathbf{w}}_k}{L_{k,j}} - {{\left| {{{\mathbf{h}}_k}{{\mathbf{w}}_k}} \right|}^2}\left(\frac{{\partial {{\bar I}_k}\left( {\mathbf{W}} \right)}}{{\partial {{\mathbf{w}}_k}}}\right)}}{{{{\left( {{L_{k,j}}} \right)}^2}}} \\
   & \qquad {+}\: \sum\limits_{l \ne k} {\frac{1}{{\ln 2}}{{\left( {1 + \frac{{{{\left| {{{\mathbf{h}}_l}{{\mathbf{w}}_l}} \right|}^2}}}{{{L_{l,m}}}}} \right)}^{ - 1}}\frac{{\left( { - 2{\mathbf{h}}_k^H{{\mathbf{h}}_k}{{\mathbf{w}}_k} - \frac{{\partial {{\bar I}_l}\left( {\mathbf{W}} \right)}}{{\partial {{\mathbf{w}}_k}}}} \right)}}{{{{\left( {{L_{l,m}}} \right)}^2}}}}
\end{split}\\
\frac{{\partial {{\bar I}_k}\left( {\mathbf{W}} \right)}}{{\partial {{\mathbf{w}}_k}}} & =  - \sum\limits_{i = 1}^K {{P_i}{r_I}\sigma _{v,i}^2\frac{{\partial {\bar {P}_{D,i}}}}{{\partial {{\mathbf{w}}_k}}}} \label{eq:partIbar}\\
&=  - \frac{{2\sigma _{f,i}^2{{\mathbf{w}}_k}}}{{\sigma _{z,i}^2}}{e^{\left( { - \frac{{{\lambda _i}}}{{\sigma _{z,i}^2}}} \right)}}\left( {\sum\limits_{r = 0} {\frac{{\lambda _i^r}}{{r!}}} \frac{{\left( {1 - r} \right)}}{{{{\left( {\sigma _{z,i}^2} \right)}^{r + 1}}}}} \right)
\end{align}
and the differential on the RHS of \eqref{eq:partIbar} is taken with respect to the average detection probability in \eqref{eq:barPDi} which is computable at the UCT.

At the $k^{th}$ iteration of the GP process, the updated precoding matrix $\W^{(k)}$ in the direction of the gradient computed above will likely not satisfy the UCT transmit power and PR interference temperature constraints. The projection step of the GP algorithm therefore projects the iterate $\W^{(k)}$ back onto the feasible constraint set $\Pwrcon$, defined as $\Pwrcon \triangleq \{ \W \mid \Tr\left( \W\W^H \right) \leq P, \Tr\left( {{\mathbf{NW}}{{\mathbf{W}}^H}{{\mathbf{N}}^H}} \right) \leq {\xi _p}\}$. Nominally, this would be achieved by determining a feasible $\W_0 \in \Pwrcon$ that is closest to $\W^{(k)}$ in terms of Frobenius norm, i.e., by minimizing the squared distance ${d^2}\left( {{{\mathbf{W}}_0},{{\mathbf{W}}^{\left( k \right)}}} \right) = \Tr\left( {{{\left( {{{\mathbf{W}}_0} - {{\mathbf{W}}^{\left( k \right)}}} \right)}^H}\left( {{{\mathbf{W}}_0} - {{\mathbf{W}}^{\left( k \right)}}} \right)} \right)$ with appropriate constraints:
\begin{subequations}
\begin{align}\label{eq:GPProject}
  \mathop {\min }\limits_{{{\mathbf{W}}_0}} & \quad {d^2}\left( {{{\mathbf{W}}_0},{{\mathbf{W}}^{\left( k \right)}}} \right) \\
  \mathrm{s.t.} &  \quad \Tr\left( \W_0\W_0^H \right) \leq P \label{eq:Projc1}\\
  {} & \quad \Tr\left( {{\mathbf{NW}_0}{{\mathbf{W}}_0^H}{{\mathbf{N}}^H}} \right) \leq {\xi _p}.\label{eq:Projc2}
\end{align}
\end{subequations}
However, instead of numerically solving the above problem, a potentially suboptimal but much simpler approach is to scale ${\mathbf{W}}^{\left( k \right)}$ such that both \eqref{eq:Projc1} and \eqref{eq:Projc2} are satisfied. This approach is partly motivated by the observation that the solution to \eqref{eq:GPProject} cannot satisfy both constraints with equality for a general channel $\mathbf{N} \ne \mathbf{I}$, and one of the constraints is guaranteed to be an inequality anyway.


A summary of the GP approach for underlay prescient sum rate maximization is
shown in Algorithm~\ref{alg_mgp}, where the step sizes $s_k$ and $\alpha_k$ are chosen using well-defined criteria such as Armijo's rule \cite[Sec. 2.3]{Bertsekas}.
\begin{algorithm}
{{\footnotesize \caption{Prescient Gradient Projection Method}} \label{alg_mgp}
\begin{algorithmic}
\STATE {\bf Initialization:} \\
\STATE Set iteration index $k=0$. \\
\STATE \quad Initialize $\W^{(0)} = [ \mathbf{w}_{1}^{(0)} \quad \mathbf{w}_{2}^{(0)} \ldots \quad \mathbf{w}_{K_u}^{(0)}]$. \\
\STATE {\bf Main Loop:} \\
\STATE \quad 1. Calculate the gradient $\nabla _{\mathbf{W}^{(k)}} \left( R_s \right)$. \\
\STATE \quad 2. Choose an appropriate step size $s_{k}$. Let $\W^{'(k)} = \W^{(k)} + s_{k}\nabla _{\mathbf{W}^{(k)}} \left( R_s \right)$\\
\STATE \quad 3. Let $\bar{\W}^{(k)}$ be the projection of $\W^{'(k)}$ onto $\Pwrcon$, where\\
\STATE \quad \quad $\Pwrcon \triangleq \{ \W \mid \Tr\left( \W \W^H \right) \leq P, \Tr\left( {{\mathbf{NW}}{{\mathbf{W}}^H}{{\mathbf{N}}^H}} \right) \leq {\xi _p}\}$. \\
\STATE \quad 4. Choose appropriate step size $\alpha_{k}$. Let $\W^{(k+1)} = \W^{(k)} +
\alpha_{k}(\bar{\W}^{(k)} - \W_{i}^{(k)})$. \\
\STATE \quad 5. $k=k+1$. If $\left\|\mbox{\rm vec}\left(\W^{(k)} - \W^{(k-1)}\right)\right\| < \epsilon$, stop; else go to step 1.
\end{algorithmic}}
\end{algorithm}

\subsection{Algorithm Based on Convex Optimization}\label{sec:ConvSDP}

While the iterative algorithm described above returns at least a
locally optimal prescient beamforming matrix, it is desirable to
investigate designs based on simpler optimization procedures.  In this
section, we investigate a suboptimal approach that maximizes the
partial underlay SINR accounting for intra-UCR and PR interference,
while making a best-effort attempt to limit the expected ICR
interference by ensuring a minimum level of signal power leakage to
them.  We first define the partial UCR SINR ${\beta _k}$ as
\begin{equation}
{\beta _k} = \frac{{\left| {{\mathbf{h}}_k {\mathbf{w}}_k } \right|^2 }}
{{\sum\nolimits_{j \ne k}^{K_u} {\left| {{\mathbf{h}}_k {\mathbf{w}}_j } \right|^2 }  + \sigma _k^2 }}, \quad k=1,\ldots,K_u,
\end{equation}
where the ICR interference term in the denominator of \eqref{eq:UCR_SINR} is omitted.
Then we pose the problem of maximizing the minimum partial UCR SINR subject to a set of constraints $\left\{ {\eta _i } \right\}_{i = 1}^K$ on the total UCT signal power received by the ICRs, as follows:
\begin{subequations}
\begin{align}
   \max\limits_{\mathbf{W}} &  \min\limits_k \beta_{k}  \\
  \mathrm{s.t.} & \Tr\left( {{\mathbf{W}} {\mathbf{W}}^H } \right) \leq P  \\
  & \Tr \left( {{\mathbf{W}} {\mathbf{W}}^H {\mathbf{F}}_i^H {\mathbf{F}}_i } \right) \geq \eta _i ,i = 1, \ldots ,K  \\
  & \Tr\left( {{\mathbf{NW}}{{\mathbf{W}}^H}{{\mathbf{N}}^H}} \right) \leq {\xi _p}
\end{align}
\end{subequations}

This can be posed as a convex optimization problem as follows. Let ${\mathbf{J}_k} \triangleq {\mathbf{w}_k}{{\mathbf{w}}_k^H}, \; \; \forall k$. 
   Applying a change of variable and relaxing the rank-1 constraints on ${\mathbf{J}_k}$, we have the reformulation
\begin{subequations}
\begin{align}\label{eq:max-minPUSINR}
  \max\limits_{\left\{ {{\mathbf{J}}_k } \right\}_{k = 1}^{K_u} } & {\text{ }}t \\
  {\mathrm{ s.t.}} \quad & t\left( {\sum\limits_{j \ne i} {\Tr \left( {{\mathbf{h}}_i^H {\mathbf{h}}_i \mathbf{J}_j } \right)}  + \sigma _i^2 } \right) - \Tr \left( {{\mathbf{h}}_k^H {\mathbf{h}}_k \mathbf{J}_k } \right) \leq 0 \label{eq:SDPSNR}\\
  & \Tr \left( {\left( {\sum\nolimits_{k = 1}^{K_u} {\sigma _{s,i}^2 {\mathbf{J}}_k } } \right){\mathbf{F}}_i^H {\mathbf{F}}_i } \right) \geq \eta _i ,{\text{ }}i = 1, \ldots ,K \\
 & \sum\nolimits_{k = 1}^{K_u} {\Tr \left( { {\mathbf{J}}_k } \right)}  \leq P \\
 & \Tr\left({{\mathbf{N}}\left(\sum\limits_{k = 1}^{K_u} {  { {\mathbf{J}}_k }} \right){{\mathbf{N}}^H}} \right) \leq {\xi _p} \\
 & t \geq 0 \\
 & {\mathbf{J}}_k  \succeq {\mathbf{0}},{\text{ }}k = 1, \ldots ,K_u.
\end{align}
\end{subequations}
In this case, however, dropping the rank constraints on $\left\{ {{\mathbf{J}}_i } \right\}_{i = i}^{K_u}$ still does not lead to a semidefinite program (SDP), since the $K_u$ underlay SNR inequality constraints in \eqref{eq:SDPSNR} are non-linear due to the fact that $t$ is a variable. Therefore, a two-stage solution strategy is required where the outer-loop performs a one-dimensional bisection search over $t$, while the inner loop solves (\ref{eq:max-minPUSINR}) for a given value of $t$, if feasible \cite{Luo08}.

\subsection{Combined Downlink and Multicast Beamforming}\label{sec:LinearComb}

As an alternative suboptimal algorithm, we present an approach with a semi-analytical expression for $\mathbf{W}$, motivated by the simple observation that the detection probability of the energy detector in~(\ref{eq:P_Dk}) increases monotonically with the received SNR at the SUs for a given false alarm rate $P_{FA,i}$.
Consider the following two extreme cases for the choice of
$\mathbf{W}$:
\begin{itemize}
\item Disregard ICRs, focus only on UCRs: If the UCT disregards the presence
of the ICRs and focuses only on its intended receivers, a
suitable choice for $\W$ is the RCI precoder
${\mathbf{W}}_{CI}$ given by~(\ref{eq:RCI}).
\item Disregard UCRs, focus only on ICRs: At this extreme, the UCT ignores
its downlink users and focuses only on improving the signal strength at the ICRs
(particularly those that could produce the most interference).  This is similar to a MIMO multicast (MC) downlink scenario, where
priority is given to certain key users.  A reasonable choice for the
transmit precoder in this case would maximize the weighted average
of the SNRs at the ICRs:
\begin{equation}\label{eq:mcb}
\mathbf{W}_{MC} = \arg\max_{\mathbf{W}}
\sum_{i=1}^K P_i N_I \sigma_{v,i}^2 \Tr \left( {\mathbf{F}}_i{{\mathbf{W}} {\mathbf{W}}^H {\mathbf{F}}_i^H  } \right) ,
\end{equation}
where the weight $P_i N_I \sigma_{v,i}^2$ measures the interference
impact of the $i$th ICR at the UCRs. The
solution to~(\ref{eq:mcb}) is given by the dominant singular
vectors of $\mathbf{F}_S^H \mathbf{\Sigma}_g \mathbf{F}_S^H$ scaled by $\sqrt{P}$, where
${\mathbf{F}}_S = \sum\nolimits_i^K \mathbf{F}_i$ and $\mathbf{\Sigma}_g$ is a
diagonal matrix with entries $P_i N_I \sigma_{v,i}^2, i=1,\cdots,K$.
\end{itemize}
Given that the prescient precoding objective is to balance
these two competing goals, a sensible approach would be to choose
$\mathbf{W}$ as some \emph{linear combination} of the solutions:
\begin{equation}\label{eq:mrtmc}
{\mathbf{W}_l} = \alpha {\mathbf{W}}_{CI}  + \left( {1 - \alpha } \right){\mathbf{W}}_{MC}
\qquad 0 \le \alpha  \le 1 \; ,
\end{equation}
where the optimal value of $\alpha \in [0,1]$ that maximizes~(\ref{eq:DirectRsMax})
can be found by a simple line search.

\section{Multi-antenna Underlay Receivers}\label{sec:MIMOUnderlayRxs}
In this section we extend the prescient downlink precoding paradigm to the case of multi-antenna UCRs with multiple data streams transmitted to each.  For simplicity, assume that each UCR is equipped with $r_u$ antennas, although the proposed algorithms hold for unequal array sizes as long as the total number of receive antennas does not exceed $t_u$. The extension to the case where the UCT serves $t_u$ spatial streams regardless of the total number of receive antennas can be made using the coordinated beamforming approach \cite{SwindleBD}, for example. The received signal at UCR $k$ is now
\begin{equation}
{{\mathbf{y}}_k} = {{\mathbf{H}}_k}{{\mathbf{W}}_k}{{\mathbf{s}}_{u,k}} + \sum\limits_{j \ne k}^{{K_u}} {{{\mathbf{H}}_k}{{\mathbf{W}}_j}{{\mathbf{s}}_{u,j}}}  + \sum\limits_{i = 1}^K {{F_i}{{\mathbf{V}}_{k,i}}{{\mathbf{s}}_{I,i}}}  + {{\mathbf{n}}_k}
\end{equation}
where $\mathbf{H}_k \in \mathbb{C}^{r_u \times t_u}$ is the main channel, ${{\mathbf{W}}_k}\in \mathbb{C}^{t_u \times l_k}$ is the beamforming matrix applied to signal $\mathbf{s}_{u,k} \in \mathbb{C}^{l_k \times 1}$ for user $k$, $F_i$ is the ICR indicator function as before, $\mathbf{s}_{I,i}$ is the $i^{th}$ ICR signal over interfering channel ${\mathbf{V}}_{k,i}\in \mathbb{C}^{r_u \times r_I}$, and ${{\mathbf{n}}_k}\sim \mathcal{CN}(\mathbf{0},\sigma_k^2 \mathbf{I})$ is additive Gaussian noise. The transmit covariance matrix for each UCR is given by ${\mathbf{Q}}_k=\W_k\W_k^H$.
We adopt a prescient block-diagonalization (PBD) strategy on the underlay downlink \cite{SwindleBD,Zhang10} to completely eliminate intra-UCR interference, as shown below.

In the first approach, the transmit covariance matrices $\{{\mathbf{Q}}_k\}^{K_u}_{s=1}$ are computed jointly so as to optimize the underlay system sum rate while subject to constraints on the PR interference and the minimum power leaked to the ICRs. The proposed PBD scheme is described mathematically as
\begin{subequations}\label{eq:JointPBD}
\begin{align}
  {}&{\max\limits_{{{\mathbf{Q}}_1}, \ldots ,{{\mathbf{Q}}_{{K_u}}}} \sum\limits_{k = 1}^{{K_u}} {{{\log }_2}\left| {{\mathbf{I}} + {{\mathbf{H}}_k}{{\mathbf{Q}}_k}{\mathbf{H}}_k^H} \right|} } \\
  \mathrm{s.t.}&\quad {{{\mathbf{H}}_k}{{\mathbf{Q}}_j}{\mathbf{H}}_k^H = {\mathbf{0}},\forall k \ne j} \\
  {}&\quad \Tr\left( {{\mathbf{N}}\left( {\sum\nolimits_{k = 1}^{{K_u}} {{{\mathbf{Q}}_k}} } \right){{\mathbf{N}}^H}} \right) \leq {\xi _p}\label{eq:PR}\\
  {}&\quad{\Tr\left({{\mathbf{F}}_i}\left( {\sum\nolimits_{k = 1}^{{K_u}} {{{\mathbf{Q}}_k}} } \right){\mathbf{F}}_i^H\right) \geq {\eta _i}} \:\forall i \label{eq:Leak}\\
  {}&\quad{\Tr \left( {\sum\nolimits_{k = 1}^{{K_u}} {{{\mathbf{Q}}_k}} } \right) \leq P} \\
{}&\quad{\mathbf{Q}}_k  \succeq {\mathbf{0}},{\text{ }}k = 1, \ldots ,K_u.\label{eq:Pwr}
\end{align}
\end{subequations}
Note that this is not equivalent to direct maximization of the UCR sum rate since the ICR interference is not included in the objective function. However, this decoupling renders the problem convex since the objective function is jointly concave and all constraints are linear in $\{\mathbf{Q}_k\}$, and the leakage constraints $\eta_i$ can be adjusted appropriately to diminish the probability of missed detections at the ICRs.

As an alternative PBD strategy, it is possible to separately design the precoding and power allocation matrices per user in a two-step process. Let
\[{{\mathbf{H}}_{ - k}} = \left[ {\begin{array}{*{20}{l}}
  {{{\mathbf{H}}_1}}& \cdots &{{{\mathbf{H}}_{k - 1}}}&{{{\mathbf{H}}_{k + 1}}}& \cdots &{{{\mathbf{H}}_{{K_u}}}}
\end{array}} \right]
\]
 represent the overall UCR downlink channel excluding the $k^{th}$ user. First, a closed-form solution for the unit-power precoding matrix of user $k$ is obtained from the nullspace of ${{\mathbf{H}}_{ - k}}$. For example, from the SVD ${{\mathbf{H}}_{ - k}} = {{\mathbf{U}}_{ - k}}{{\mathbf{\Sigma }}_{ - k}}{\left[ {\begin{array}{*{20}{c}}
  {{{\mathbf{V}}_{ - k,1}}}&{{{\mathbf{V}}_{ - k,0}}}
\end{array}} \right]^H}$, the last $(t_u-l_k)$ right singular vectors contained in ${\mathbf{V}}_{ - k,0}$ can be used to construct $\mathbf{W}_k$ \cite{SwindleBD}. However, unlike the conventional BD algorithm, the power allocated over the $l_k$ spatial modes of user $k$ is now no longer obtained via waterfilling. Let $\rank({{\mathbf{H}}_k}{{\mathbf{W}}_k})=r_k$ for user $k$'s effective channel, and assume $l_k=r_k$. Consider the SVD of user $k$'s effective channel ${{\mathbf{H}}_k}{{\mathbf{W}}_k} = {{\mathbf{U}}_k}{{\mathbf{\Sigma }}_k}{\mathbf{V}}_k^H$ where ${{\mathbf{\Sigma }}_k}=\operatorname{diag}\left(\epsilon_{k,1},\ldots,\epsilon_{k,r_k}\right)$ is a $r_k \times r_k$ diagonal matrix, and define $\mathbf{\Lambda}_k=\operatorname{diag}\left(\lambda_{k,1},\ldots,\lambda_{k,r_k}\right)$ to be the power allocation matrix. The overall downlink power allocation matrix is therefore ${{\mathbf{\Lambda }}_u} = \operatorname{blkdiag} \left( {{{\mathbf{\Lambda }}_1}, \ldots ,{{\mathbf{\Lambda }}_{{K_u}}}} \right)$. The PR interference and ICR signal power constraints are accommodated in the power allocation step based on a numerical optimization:
\begin{subequations}\label{eq:SepPBD}
\begin{align}
  {}&\mathop {\max }\limits_{\mathbf{\Lambda}_u} \sum\limits_{k = 1}^{{K_u}} {\sum\limits_{m = 1}^{{l_k}} {{{\log }_2}\left( {1 + \varepsilon _{k,m}^2{\lambda _{k,m}}} \right)} }  \hfill \\
  \mathrm{s.t.}&\quad\sum\nolimits_{c=1}^{r_p}\sum\nolimits_{k=1}^{K_u} {\sum\nolimits_{m=1}^{l_k} {\left\| {{{\mathbf{n}}_{c}}{{\mathbf{w}}_{k,m}}} \right\|_2^2{\lambda _{k,m}}} }  \geq {\xi_p} \hfill \label{eq:reformPR}\\
  {}&\quad\sum\nolimits_{n=1}^{r_I}\sum\nolimits_{k=1}^{K_u} {\sum\nolimits_{m=1}^{l_k} {\left\| {{{\mathbf{f}}_{i,n}}{{\mathbf{w}}_{k,m}}} \right\|_2^2{\lambda _{k,m}}} }  \geq {\eta _i}, i=1,\ldots,K, \label{eq:reformLeak}\hfill \\
  {}&\quad\sum\nolimits_{k=1}^{K_u} {\sum\nolimits_{m=1}^{l_k} {\left\| {{{\mathbf{w}}_{k,m}}} \right\|_2^2{\lambda _{k,m}}} }  \leq P \label{eq:reformPwr}\hfill
\end{align}
\end{subequations}
where ${\mathbf{n}}_{c}$ is the $c^{th}$ row of $\mathbf{N}$,  ${\mathbf{f}}_{i,n}$ is the $n^{th}$ row of $\mathbf{F}_i$, and ${\mathbf{w}}_{k,m}$ is the $m^{th}$ column of $\mathbf{W}_k$. The leakage and power constraints (\ref{eq:reformPR})-(\ref{eq:reformPwr}) are equivalent to (\ref{eq:PR})-(\ref{eq:Pwr}). This is a convex program since the objective function is concave and all constraints are linear in $\{\lambda_{k,i}\}$, and can be efficiently solved using interior-point methods.
It must be noted however that a separate design of the underlay precoding and power allocation matrices is potentially suboptimal compared to the joint design of (\ref{eq:JointPBD}).

\section{Simulation Results}\label{sec:Sim}
In this section, we present the results of several numerical
experiments to verify the improvement in primary link performance provided by
prescient beamforming. To avoid repetition, unless specified
otherwise, all results in this section are based on the partial CSI
model with instantaneous CSI of the downlink and UCT-ICR links, and only
statistical CSI of the ICR-to-UCR links available to the underlay
transmitter. Each channel realization for all terminals is drawn from
a zero-mean circularly symmetric complex Gaussian distribution, and
all results are averaged over 1000 channel realizations. The background AWGN
variance at all receivers is assumed to be unity, the primary antenna array sizes are fixed as $t_p=r_p=4$ with PT transmit power $P_t=10$, and the PR interference cap is set to $\xi_p=10$. The
convex programs are solved numerically using the {\sf{cvx}} MATLAB
toolbox \cite{cvx}. At the ICRs we set the transmit power to $P_i=20dB$, false alarm rate target $P_{FA,i}=10^{-3}\; \forall i$, and sample size of
$\tilde M=4$. The prescient GP algorithm is
run 5 times for each set of channel realizations with four random
initializations and an initialization based on the na\"{\i}ve RCI precoder to reduce the likelihood of a local maximum; the
best-performing precoding solution is chosen as the result.

\begin{figure}[htbp]
\centering
\includegraphics[width=\linewidth]{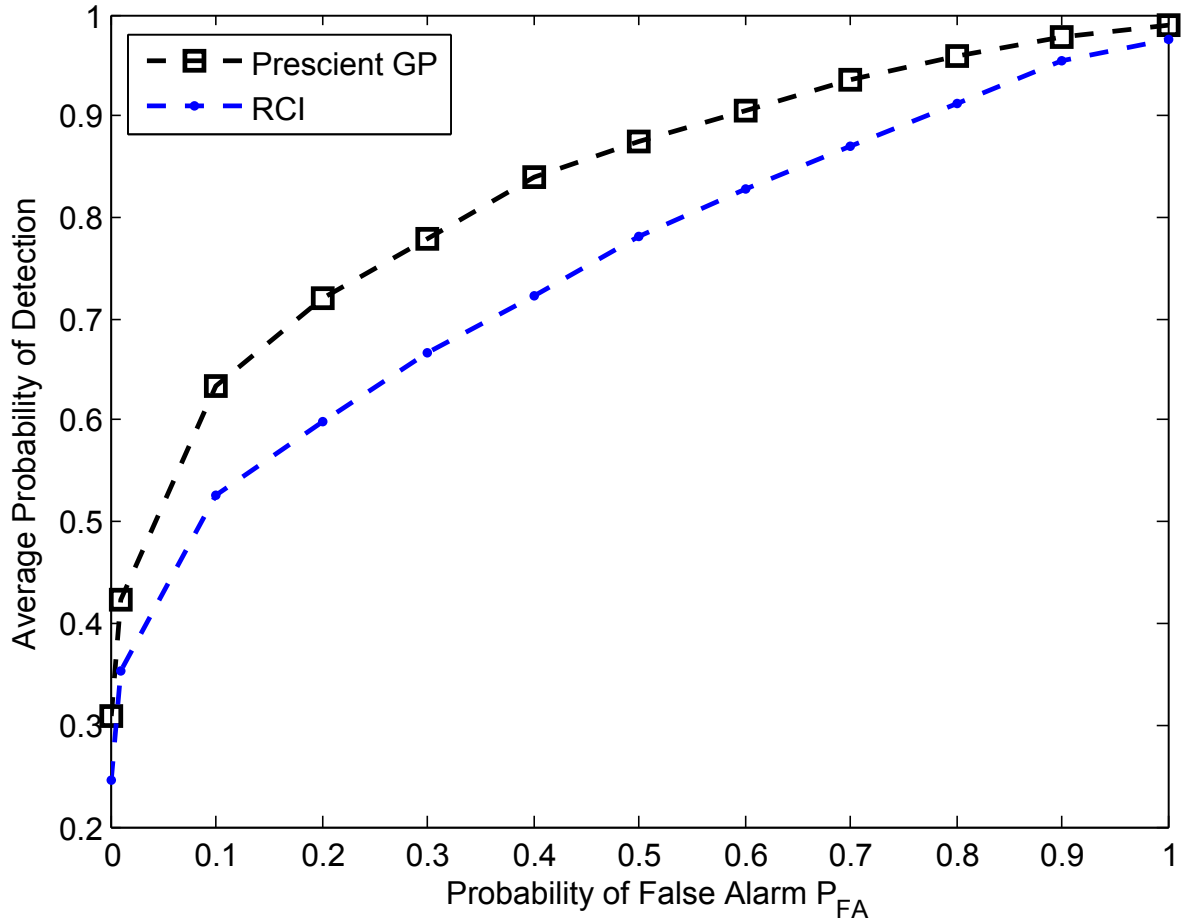}
\caption{ROC curve for energy detection comparing prescient precoding with regularized channel inversion, $t_u=K_u=3,r_I=K=2,P=15dB$.}
\label{fig_ROC}
\end{figure}

In Fig.~\ref{fig_ROC}, we first examine the energy detection
rec\-eiv\-er-op\-erat\-ing-char\-act\-er\-is\-tic at an arbitrary ICR for prescient GP precoding compared to RCI transmission for $K_u=3$ single-antenna UCRs.
The UCT transmit power is fixed at $P=10dB$ with $t_u=3$ antennas, and $K=2$ ICRs are present with $r_I=2$ antennas each.  We
observe that prescient precoding provides a significant improvement in
energy detection performance for the entire range of $P_{FA}$, and consequently reduces the likelihood of ICR missed detections.

\begin{figure}[htbp]
\centering
\includegraphics[width=\linewidth]{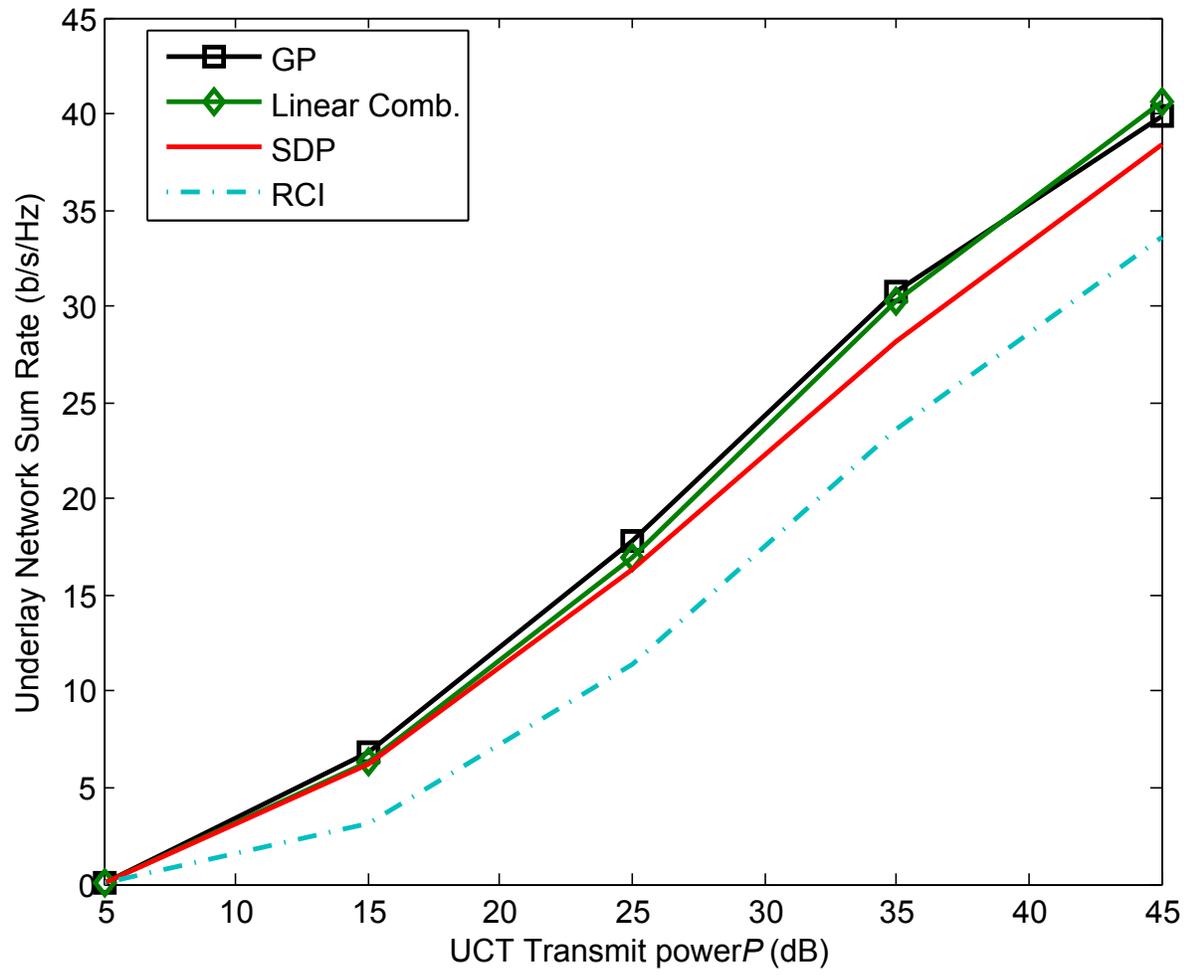}
\caption{Underlay sum rate for prescient algorithms and RCI precoding with $t_u=K_u=K=3,r_I=2$.}
\label{fig_SumRvsP}
\end{figure}

Sum rate results for the single-antenna UCR downlink versus UCT transmit power with
$t_u=K_u=K=3,r_I=2$ are shown in Fig.~\ref{fig_SumRvsP}. The prescient schemes improve markedly
upon the na\"{\i}ve RCI precoder since each ICR with a
missed detection interferes with multiple UCRs. The linear combination scheme is observed to be a very competitive alternative compared to the computationally intensive GP solution. The SDP-based prescient scheme suffers from the difficulty of optimally choosing leakage power thresholds $\eta_i$. The
proposed prescient GP precoder provides an increase of up to 7
(bits/s/Hz) in spectral efficiency compared to the RCI scheme, which highlights the significant benefit of
preemptively mitigating secondary user interference.

\begin{figure}[htbp]
\centering
\includegraphics[width=\linewidth]{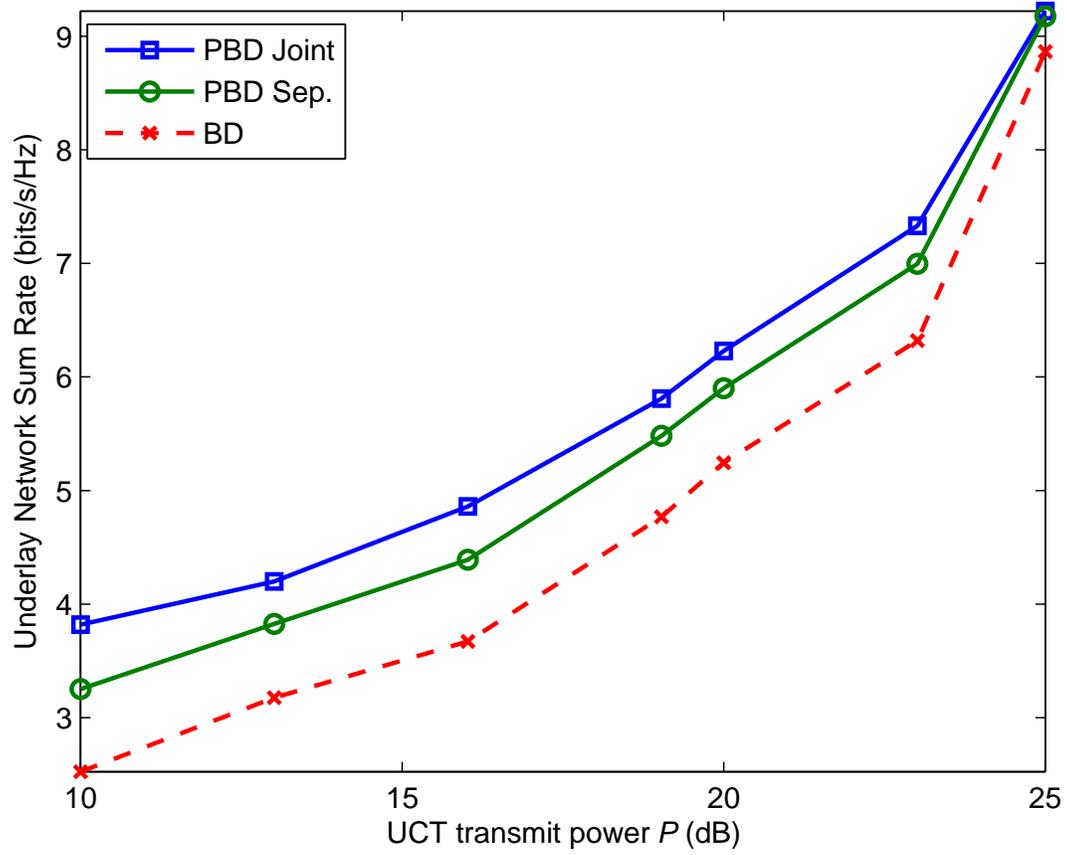}
\caption[Underlay sum rate for prescient and conventional block-diagonalization.]{Underlay sum rate for prescient and conventional block-diagonalization with $t_u=8,K_u=4,K=r_I=r_u=2$.}
\label{fig_PBDSumRvsP}
\end{figure}

We now consider prescient versus conventional block-diagonalization schemes  for the multi-antenna UCR downlink with $r_u=2$. In Fig.~\ref{fig_PBDSumRvsP} the underlay sum rate is displayed as a function of the UCT transmit power for $t_u=8,K_u=4,K=r_I=2$. The greatest benefit of the PBD schemes is observed at low to intermediate SNRs, while the sum rate of all three algorithms gradually converge at high SNR. This is because the diversion of transmit power to the ICRs under PBD now has a greater penalty in terms of spatial multiplexing loss to the multi-antenna UCRs.

\begin{figure}[htbp]
\centering
\includegraphics[width=\linewidth]{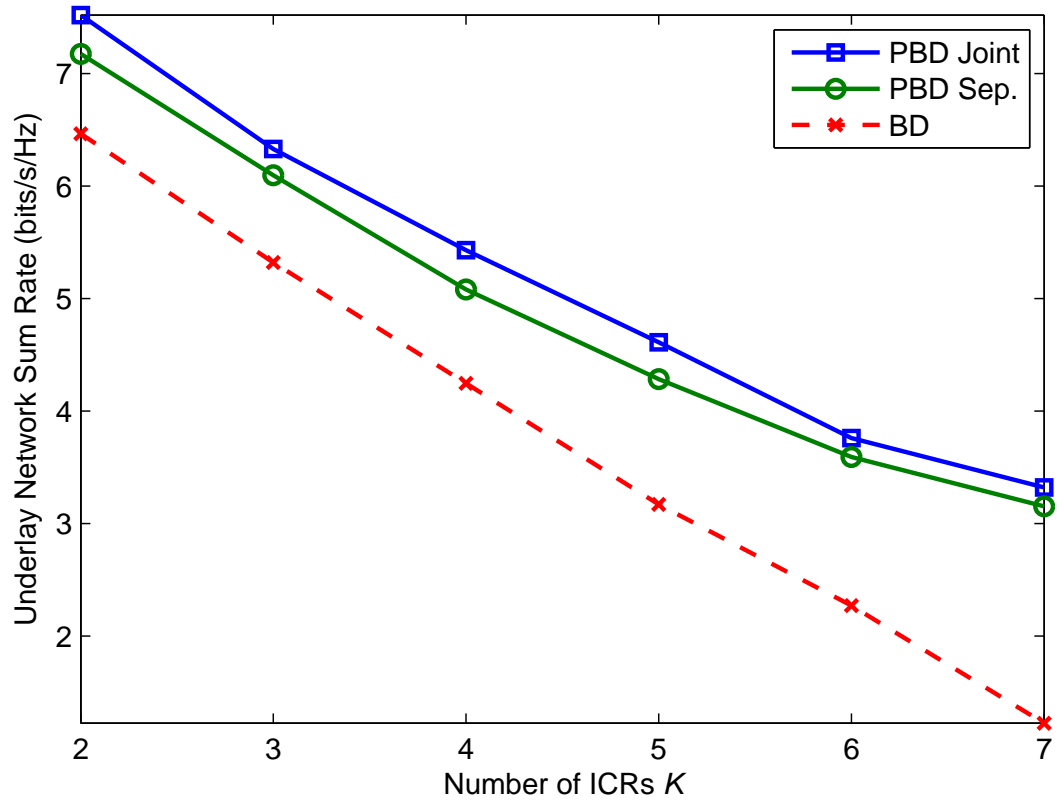}
\caption{Underlay sum rate for prescient and conventional block-diagonalization versus number of ICRs $K$, for $t_u=6,K_u=3,r_I=r_u=2,P=100$.}
\label{fig_PBDSumRvsK}
\end{figure}

Finally, Fig.~\ref{fig_PBDSumRvsK} presents the PBD and BD underlay sum rates as the number of potentially interfering ICRs increases, for fixed UCT power $P=15dB$. The relatively low combined transmit power of the UCT and PT leads to a potentially significant number of missed detections at the ICRs, and the expected ICR interference clearly worsens as $K$ increases. This is especially true for the conventional BD scheme, which suffers from a pronounced degradation in sum rate since it neglects the sensing performance of the ICRs. An important implication of this outcome is that the successful coexistence of UCRs and ICRs in a heterogeneous DSA network cannot be assured merely by modifying the UCT precoding strategy; smarter ICR spectrum sensing approaches must also be adopted.

\section{Conclusion}\label{sec:Conclusion}
This work has examined a novel heterogeneous DSA network where the
primary users coexist with both underlay and interweave cognitive
radios, all terminals being potentially equipped with multiple
antennas. We investigated the design of MIMO precoding algorithms and
the underlay transmitter in order to increase the detection
probability at the ICRs, while simultaneously meeting a desired
Quality-of-Service target for the underlay receivers and constraining
the amount of interference leaked to the PUs. The objective of such a
proactive approach, referred to as prescient precoding, is to minimize
the probability of interference from ICRs to the UCR and PU receivers
due to imperfect spectrum sensing. We presented three different
downlink prescient precoding algorithms for the case of multiple
single-antenna UCRs and multi-antenna PUs/ICRs.  We then presented
prescient block-diagonalization algorithms for the MIMO underlay
downlink where spatial multiplexing is performed for multiple
multi-antenna UCR receivers.  Numerical experiments demonstrate that
prescient precoding by the UCT preemptively mitigates missed
detections at the ICRs, and provides a significantly pronounced
performance gain in underlay sum rate compared to conventional
precoding strategies. For future study, it is of interest to design
prescient user selection algorithms to accommodate scenarios where the
number of UCRs exceeds the number of UCT transmit antennas.

\linespread{0.99}

\newpage

\end{document}